\documentclass{jetpl}
\twocolumn

\lat

\title{ Primordial pairing and binding of superheavy charged 
particles in the Early Universe.}

\rtitle{ Primordial pairing and binding \ldots}

\sodtitle{ Primordial pairing and binding of superheavy charged 
particles in the Early Universe}

\author{V.\,K.\,Dubrovich$^{+}$ \/\thanks{e-mail:dubr@MD1381.spb.edu},
M.\,Yu.\,Khlopov$^{*,**,***}$}

\rauthor{V.\,K.\,Dubrovich V.K., M.\,Yu.\,Khlopov M.Yu.}

\sodauthor{Dubrovich, Khlopov}

\address{$^+$Special Astrophysical Observatory RAS,
196140 St.Petersburg, Russia\\~\\
$^{*}$M.V.Keldysh Institute of Applied Mathematics RAS, 
125047, Moscow, Russia.\\~\\
$^{**}$Moscow Engineering Physics Institute (Technical University), 
Moscow, Russia.\\~\\
$^{***}$Physics Department, Universita' degli studi "La Sapienza",5,
Piazzale Aldo Moro - I 00185 Roma, Italy.}

\dates{*}{*}

\abstract{ Primordial superheavy particles, considered as the source of 
Ultra High Energy Cosmic Rays (UHECR) and produced in local processes in the
early Universe, should bear some strictly or approximately
conserved charge to be sufficiently stable to survive to the present time.
Charge conservation makes them to be produced in pairs,
and the estimated separation of particle and antiparticle in such pair
is shown to be in some cases much smaller than the average separation 
determined by the averaged number density of considered particles.
If the new U(1) charge is the source of a long range field similar to
electromagnetic field, the particle and antiparticle, possessing that charge, 
can form primordial bound 
system with annihilation timescale, which can satisfy the conditions,
assumed for this type of UHECR sources. These conditions severely constrain 
the possible properties of considered particles.
}

\PACS{*}

\begin{document} 

\maketitle  

The origin of cosmic rays with energies, exceeding the GZK cut off
energy ~[1], is widely discussed, and one of popular approaches
is related with decays or annihilation in the Galaxy of primordial
superheavy particles [2\ch 5] (see ~[5] for review and references where in).
The mass of such particles is assumed to be higher than the reheating
temperature of inflationary Universe, so it is assumed that the particles
are created in some nonequilibirium processes, such as inflaton decay ~[6]
at the stage of preheating after inflation ~[7].

The problems, related with this approach, are as follows. If the source of
ultra high energy cosmic rays (UHECR) is related with particle decay in
the Galaxy, the timescale scale of this decay, which is necessary
to reproduce the UHECR data, needs special nontrivial explanation. Indeed,
the relic unstable particle should survive to the present time, and having
the mass $m$ of the order of $10^{14}$ GeV or larger it should have the
lifetime $\tau$,
exceeding the age of the Universe. On the other hand, even, if particle decay
is due to gravitational interaction, and its probability is of the order of 
$\frac{1}{\tau}=(\frac{m}{m_{Pl}})^{4} m$,
where $m_{Pl}=10^{19}$ GeV is the Planck mass, the estimated lifetime would be
by many orders of magnitude smaller. It implies some specific additional
suppression factors in the probability of decay, which need rather nontrivial
physical realisation (~[2], ~[5]), e.g. in the model of cryptons ~[8] 
(see ~[9] for review).

If the considered particles are absolutely stable,  the source
of UHECRs is related with their annihilation in the Galaxy. But their
averaged number density, constrained by the upper limit on their total density,
is so low, that strongly inhomogeneous distribution is needed to enhance the
effect of annihilation to the level, desired to explain the origin of UHECR by
this mechanism.

In the present note, we offer new approach to the solution of
the latter problem. If superheavy particles possess new U(1) gauge charge,
related to the hidden sector of particle theory, they are created in pairs.
The  Coulomb-like attraction (mediated by the massless U(1) gauge boson) 
between particles and antiparticles in these pairs
can lead to their primordial binding, so that the annihilation
in the bound system provides the mechanism for UHECR origin.

Note, first of all, that in quantum theory particle stability reflects
the conservation law, which according to Noether's theorem is related
with the existence of a conserved charge, possessed by the considered
particle. Charge conservation implies that particle should be created
together with its antiparticle. It means that, being stable,
the considered superheavy
particles should bear a conserved charge, and
such charged particles should be created in pairs with their antiparticles
at the stage of preheating.

Being created in the local process of
inflaton field decay the pair is localised within the cosmological
horizon in the period of creation. If the momentum distribution
of created particles is peaked below $p \sim mc$, they don't
spread beyond the proper region of their original localization,
being in the period of creation $l \sim c/H$, where
the Hubble constant $H$ at the preheating stage
is in the range $H_{r} \le H \le H_{end}$.
Here $H_{end}$ is the Hubble constant in the end of inflation
and $H_{r}$ is the Hubble constant in the period of reheating.
For relativistic pairs the region of localization is determined
by the size of cosmological horizon in the period
of their derelativization.
In the course of successive expansion the distance $l$ between
particles and antiparticles grows with the scale factor,
so that after reheating at the temperature $T$ it is
equal to (here and further, if not indicated otherwise,
we use the units $\hbar = c = k = 1$)
\begin{equation}
l(T) = (\frac{m_{Pl}}{H})^{1/2} \frac{1}{T}.
\end{equation}

The averaged number density of superheavy particles $n$
is constrained by the upper limit on their modern density.
Say, if we take their maximal possible contribution in the units of critical 
density, $\Omega_{X}$, not to exceed 0.3, the modern cosmological average number
density should be $n = 10^{-20}\frac{10^{14} GeV}{m} \frac{\Omega_{X}}{0.3}cm^{-3}$
(being $n= 4 \cdot 10^{-22} \frac{10^{14}GeV}{m} \frac{\Omega_{X}}{0.3} T^{3}$ in
the units $\hbar = c = k = 1$ at the temperature $T$). 
It corresponds
at the temperature $T$ to the mean distance ($l_{s} \sim n^{-1/3}$)
equal to  
\begin{equation}
l_{s} \approx 1.6 \cdot 10^{7} (\frac{m}{10^{14}GeV})^{1/3}(\frac{0.3}{\Omega_{X}})^{1/3} \frac{1}{T}.
\end{equation}

One finds that superheavy nonrelativistic particles,
created just after the end of inflation,
when $H \sim H_{end} \sim 10^{13} GeV$, are separated from their
antiparticles at distances more than 4 orders of magnitude
smaller, than the average distance between these pairs. On the other hand,
if the nonequilibrium processes of superheavy particles creation 
(such as decay of inflaton) take place in the end of preheating stage,
and the reheating temperature is as low as it is constrained from the effects 
of gravitino decays on $^{6}Li$ abundance ($T_{r}< 4\cdot 10^{6}$GeV ~[10,~11]),
the primordial separation of pairs, given by Eq(1), can even exceed 
the value, given by Eq.(2).
It means that the separation between particles and antiparticles can be determined 
in this case by their 
averaged density, if they were created at $H \le H_{s}\sim 10^{-15} \cdot m_{Pl}
(\frac{10^{14}GeV}{m})^{2/3}(\frac{\Omega_{X}}{0.3})^{2/3} \sim 10^{4}(\frac{\Omega_{X}}{0.3})^{2/3}$ GeV.

If the considered charge is the source of a long range field,
similar to the electromagnetic field, which can bind
particle and antiparticle into the atom-like system,
analogous to positronium, it may have important practical 
implications for UHECR problem. The annihilation timescale
of such bound system can provide the rate of UHE particle sources,
corresponding to UHECR data.

The pair of particle and antiparticle with opposite gauge charges forms bound system,
when in the course of expansion the absolute magnitude of potential energy of pair $V= \frac{\alpha_{y}}{l}$
exceeds the kinetic energy of particle relative motion $T_{k}= \frac{p^{2}}{2m}$.
The mechanism is similar to the proposed in ~[12]
for binding of magnetic monopole-antimonopole pairs. It is not a
recombination one. The binding of two opositely charged particles
is caused just by their Coulomb-like attraction, once it exceeds the kinetic energy
of their relative motion.

In case, plasma interactions do not heat superheavy particles,
created with relative momentum $p \le mc$ in the period,
corresponding to Hubble constant $H \ge H_{s}$, their initial separation,
being of the order of 
\begin{equation}
l(H) = (\frac{p}{mH}),
\end{equation}
experiences only the effect of general expansion, proportional to the inverse
first power of the scale factor, while the initial kinetic energy decreases
as the square of the scale factor. Thus, the binding condition is fulfilled 
in the period, corresponding to the Hubble constant $H_{c}$, determined by
the equation
\begin{equation}
(\frac{H}{H_{c}})^{1/2} = \frac{p^{3}}{2 m^{2}\alpha_{y}H},
\end{equation}
where $H$ is the Hubble constant in the period of particle creation and $\alpha_{y}$ 
is the "running constant" of the long range U(1) interaction,
possessed by the superheavy particles. If the local process of pair creation does not involve nonzero orbital momentum, due to the 
primordial pairing the bound system is formed in the state with zero orbital momentum.
The size of bound system
strongly depends on the initial momentum distribution and for $p \le mc$ equals
\begin{equation} 
l_{c} = \frac{p^{4}}{2 \alpha_{y} m^{3} H^{2}} = 2 \frac{\alpha_{y}}{m \beta^{2}},
\end{equation}
where
\begin{equation} 
\beta = \frac{2 \alpha_{y} m H}{p^{2}}.
\end{equation}
The annihilation timescale of this bound system can
be estimated from the annihilation rate, given by
\begin{equation}
w_{ann} = \mid \Psi (0) \mid ^{2} (\sigma v)_{ann} \sim l_{c}^{-3}
\frac{\alpha_{y} ^{2}}{m^{2}} C_{y},
\end{equation}
where the "Coulomb" factor $C_{y}$ arises similar to 
the case of a pair of electrically charged particle and antiparticle.
For the relative velocity $\frac{v}{c}\ll 1$ it is given by [13]
$C_{y} = \frac{2 \pi \alpha_{y} c}{v}$. 
Finally, taking $v/c \sim \beta$, one obtains for the annihilation timescale
\begin{equation}
\tau_{ann} \sim \frac{1}{8 \pi \alpha_{y}^{5}} (\frac{p}{mc})^{10} 
(\frac{m}{H})^{5} \frac{1}{m} = \frac{4}{\pi m \beta^{5}}.
\end{equation}

For $H_{end}\ge H \ge H_{s}$, the
annihilation timescale equals
\begin{equation}
\tau_{ann} =2\cdot 10^{19} (\frac{1}{50 \alpha_{y}})^{5} (\frac{p}{mc})^{10} (\frac{10^{4}GeV}{H})^{5} (\frac{m}{10^{14}GeV})^{4} s,
\end{equation}
being for $p \sim mc$, $\alpha_{y}= \frac{1}{50}$ 
and $m = 10^{14}$ GeV in the range from $10^{-26}$s
up to $2 \cdot 10^{19}(\frac{0.3}{\Omega_{X}})^{10/3}$s.
The size of a bound system is given by 
\begin{equation}
l_{c} = 5 \cdot 10^{-7} (\frac{p}{mc})^{4} \frac{m}{10^{14}GeV} 
(\frac{10^{4}GeV}{H})^{2} cm,
\end{equation}
ranging for $2 \cdot 10^{-10} \le \frac{\Omega_{X}}{0.3} \le 1$
from $7 \cdot 10^{-7}$cm to $6 \cdot 10^{-3}$cm.

Provided that the primordial abundance of superheavy particles,
created on preheating stage corresponds to the appropriate modern density
$\Omega_{X} \le 0.3$, and the annihilation timescale exceeds the age of the Universe
$t_{U} = 4 \cdot 10^{17}$s, owing to strong dependence on initial momentum $p$, the magnitude
$r_{X} = \frac{\Omega_{X}}{0.3} \frac{t_{U}}{\tau_{X}}$
can reach the value $r_{X} = 2 \cdot 10^{-10}$,
which was found in ~[2] to fit the UHECR data by superheavy
particle decays in the halo of our Galaxy.

In the case of late particle production (i.e. at $H \le H_{s}$) the binding condition 
can retain the form (4), if $l(H) \le l_{s}$.
In the opposite case, when $l(H) \ge l_{s}$,
the primordial pairing is lost and even being produced
with zero orbital momentum particles and antiparticles, originated from different pairs, 
in general, form bound systems with nonzero orbital momentum.
The size of the bound system is in this case obtained from the binding condition
for the initial separation, determined by Eq.(2), and it is equal to
\begin{equation}
l_{c} \approx \frac{10^{15}}{2 \alpha_{y} m_{Pl}}(\frac{m}{10^{14}GeV})^{2/3} (\frac{0.3}{\Omega_{X}})^{2/3} (\frac{p}{mc})^{2}(\frac{m}{H}).
\end{equation}
The lifetime of the bound system can be reasonably estimated in this case with the use of the well known results of classical problem
of the falling down the center due to radiation in the bound system of massive particles with opposite electric charges.
The corresponding timescale is given by (see [14] for details)
\begin{equation}
\tau =  \frac{l_{c}^{3}}{64 \pi} \frac{m^{2}}{\alpha_{y}^{2}}. 
\end{equation}
Using the Eq.(11) and the condition $l(H) \ge l_{s}$, one obtains for this case
the following restriction
\begin{equation}
r_{X} = \frac{\Omega_{X}}{0.3} \frac{t_{U}}{\tau_{X}} \le 3 \cdot 10^{-10} (\frac{\Omega_{X}}{0.3})^{5} (\frac{10^{14}GeV}{m})^{9}.
\end{equation}

The gauge U(1) nature of the charge, possessed by superheavy particles, assumes the existence of massless U(1) gauge bosons
(y-photons) mediating this interaction. Since the considered superheavy particles are the lightest particles
bearing this charge, and they are not in thermodynamical equilibrium, one can expect that there should be no thermal background
of y-photons and that their non equilibrium fluxes can not heat significantly the superheavy particles.

The situation changes drastically, if the superheavy particles  possess not only new U(1) charge but also
some ordinary (weak, strong or electric) charge.
Due to this charge superheavy particles interact with the equilibrium relativistic plasma
(with the number density $n \sim T^{3}$)
and for the mass of particles $m \le \alpha^{2} m_{Pl}$ the rate of heating
$n \sigma v \Delta E \sim \alpha^{2} \frac{T^{3}}{m}$
is sufficiently high 
to bring the particles into thermal 
equilibrium with this plasma.
Here $\alpha$ is the running constant of
the considered (weak, strong or electromagnetic) interaction. 

Plasma heating
causes the thermal motion of superheavy particles. 
At $T \le m (\frac{m}{\alpha^{2} m_{Pl}})^{2}$ their mean free path 
relative to scattering with plasma exceeds the free thermal motion path,
so it is not diffusion, but free motion with thermal velocity $v_{T}$ that leads to 
complete loss of initial pairing, since  $v_{T}t$ formally exceeds $l_{s}$
at $T \le 10^{-10} m_{Pl}(\frac{\Omega_{X}}{0.3})^{2/3} (\frac{10^{14}GeV}{m})^{5/3}$.

While plasma heating keeps superheavy particles
in thermal equilibium the binding condition $V \ge T_{kin}$
can not take place. At $T < T_{N}$, (where 
$N = e, QCD, w$ respectively, and  $T_{e} \sim 100$keV for electrically charged particles;
$T_{QCD} \sim 300$MeV for coloured particles and $T_{w} \approx 20$GeV for weakly interacting particles,
see [14] for details)
the plasma heating is suppressed and superheavy particles go out
of thermal equilibrium.

In the course of successive expansion the binding
condition is formally reached at $T_{c}$, given by
\begin{equation}
T_{c} = T_{N} \alpha_{y} 3 \cdot 10^{-8} (\frac{\Omega_{X}}{0.3})^{1/3} (\frac{10^{14}GeV}{m})^{1/3}.
\end{equation} 
However, for electrically charged particles, 
the binding in fact does not take place to the present time, since  one gets from Eq. (14) $T_{c} \le 1$K.
Bound systems of hadronic and weakly interacting superheavy particles can form, respectively,
at $T_{c} \sim 0.3$eV and $T_{c} \approx 20$eV, but  even for weakly interacting particles 
the size of such bound systems approaches a half of meter (30 m for hadronic particles!). 
It leads to extremely long annihilation timescale of
these bound systems, that can not fit UHECR data. 
It makes impossible to realise the considered mechanism of UHECR origin, 
if the superheavy U(1) charged particles share
ordinary weak, strong or electromagnetic interactions. 

Disruption of primordial bound systems in their collisions and by tidal forces in the Galaxy
reduces their concentration in the regions of enhanced density. 
Such spatial distribution, specific for these UHECR
sources, makes possible to distinguish
them from other possible  mechanisms ~[4, ~9, ~15] in the future AUGER and EUSO experiments.

The crucial physical condition for the
formation of primordial bound systems of superheavy particles is the existence
of new strictly conserved local U(1) gauge symmetry, ascribed to the hidden sector of
particle theory. Such symmetry can arise in the extended variants of GUT 
models (see e.g. ~[10] for review), in heterotic string phenomenology (see ~[13] and references wherein)
and in D-brane phenomenology ~[16]. Note, that in such models the strictly conserved SU(2) symmetry can
also arise in the hidden sector, what leads to a nontrivial mechanism of primordial binding of superheavy 
particles due to macroscopic size SU(2) confinement, as it was the case for "tetons" ~[17]. 

The proposed mechanism offers the link between the observed UHECRs
and the predictions of particle theory, which can not be tested by any other means
and on which the analysis of primordial pairing and binding can put severe constraints. 
If viable, the considered mechanism
makes UHECR the unique source of detailed information on the possible
properties of the hidden sector of particle theory and on the
physics of very early Universe.

The work was partially performed in the framework
of State Contract 40.022.1.1.1106
and supported in part by RFBR grant 02-02-17490
and grant UR.02.01.026. One of us (MYuKh) expresses his gratitude to
IHES, LUTH (Observatory Paris-Meudon) for hospitality and the both
authors are grateful for support of their visit to College de France
(Paris, April 2002). We are grateful to D.Fargion and S.Sarkar for
useful comments.

\end{document}